\UseRawInputEncoding

 \documentclass[final,5p,times,twocolumn,authoryear]{elsarticle}

\usepackage{amssymb}
\usepackage{lipsum}
\usepackage{amsmath}
\usepackage{natbib}
\usepackage{xcolor}
\usepackage{graphics}

\journal{Physics Letters B}

\begin{document}

\begin{frontmatter}



\title{Accretion of the degenerate Fermi gas onto a Reissner-Nordstr\"{o}m black hole}


\author[first,second]{Ping Li}
\ead{lip57120@huas.edu.cn}
\affiliation[first]{organization={College of Mathematics and Physics,Hunan University of Arts and Sciences},
            addressline={3150 Dongting Dadao}, 
            city={Changde City},
            postcode={415000}, 
            state={Hunan Province},
            country={China}}
\affiliation[second]{organization={Hunan Province Key Laboratory Integration and Optical Manufacturing Technology},
            addressline={3150 Dongting Dadao}, 
            city={Changde City},
            postcode={415000}, 
            state={Hunan Province},
            country={China}}

\author[first,third]{Jiang-he Yang}
\affiliation[third]{organization={Center for Astrophysics, Guangzhou University},
            addressline={230 West Ring Road}, 
            city={Guangzhou},
            postcode={510006}, 
            state={Guangdong Province},
            country={China}}

\author[first,second]{Siwei Xu}

\begin{abstract}
We extend the Rioseco and Sarbach model into Reissner-Nordstr\"{o}m black hole accretes degenerate relativistic Fermi gas. The accretion theory is based on the Boyer-Lindquist coordinates and the Fermi gas follows Fermi-Dirac statistics at infinity. The expressions for the particle current density, the stress energy-momentum tensor, and three accretion rates are derived. We first investigated the impact of Risoseco and Sarbach model on the evolution of a black hole's charge. The results show that both the mass accretion rate and charge accretion rate are proportional to the particle accretion rate. We have also provided analytical results at infinity and numerical results within a finite range for these quantities. Our results indicate that the accretion rate decreases as the charge of the black hole increases, suggesting that our accretion model does not violate the cosmic censorship hypothesis. In this paper, we first point out that the accretion of Vlasov gas behaves as an anisotropic fluid containing two perfect-fluid components. One component represents the isotropic fluid of Fermi gas, while the other represents a null fluid. When using the Boyer-Lindquist coordinate system, we observed that the contribution from the null fluid persists even at infinity, which led to the radial pressure always smaller than the tangential pressure. Therefore, it's not appropriate to treat the accretion model as a perfect fluid at infinity.
\end{abstract}




\end{frontmatter}




\section{Introduction}
In 2017, Rioseco and Sarbach proposed an accretion model \citep{Rioseco2017-1} for Schwarzschild black holes that gained considerable attention. They assumed that the gas, composed of a large number of particles, is in local thermodynamic equilibrium at infinity, with rare collisions between particles and negligible self-gravity, also known as the collisionless Vlasov gas. The conditions of rare collisions was supported by observations from the Event Horizon Telescope Collaboration \citep{EHTL1,EHTL5,EHTL12,EHTL16}. As the particles travel along the geodesic into finite range, they form steady-state, spherically symmetric accretion flows. The researchers solved the relativistic Vlasov equation in Schwarzschild geometry and obtained the condition satisfied by the distribution function with an appropriate symmetry. In the case of the Maxwell-J\"{u}ttner distribution, they derived the current density, the stress energy-momentum tensor, and the mass accretion rate. In another paper \citep{Rioseco2017-2}, they also performed numerical computations of these quantities and found that the radial pressure and tangential pressure approximately equal each other at infinity, suggesting the accretion model can be treated as an isotropic perfect fluid. However, as an observer approaches the horizon, the tangential pressure increases steadily while the radial pressure decreases. At the horizon, the tangential pressure becomes larger than the radial pressure. This provides a partial explanation for the low accretion rate problem, also known as the Bondi-Hoyle-Lyttleton accretion problem \citep{Hoyle1939,Lyttleton1940,Bondi1944,Chaverra2015}.

Since then, many studies have expanded this model to encompass more general scenarios. The Rioseco and Sarbach's accretion model for the Reissner-Nordstr\"{o}m black hole were generalized by Ci\'{s}slik and Mach \citep{Cieslik2020}. While in Refs. \citep{Mach2021-1,Mach2021-2,Mach2022}, authors delved into the accretion of relativistic Vlasov gas onto a moving Schwarzschild black hole. The results are using to model the growth of primordial black holes in the early Universe. Gamboa \textit{et. al.} also explored the accretion of Vlasov gas onto a Schwarzschild black hole at a sphere of finite radius \citep{Gamboa2021}. Additionally, beyond static spherical black holes in general relativity, Rioseco and Sarbach's results were extended to black holes in modified gravities. Liao and Liu looked into the accretion of collisionless Vlasov gas onto a Bardeen regular black hole \citep{Liao2022}, while Cai and Yang studied the accretion of Vlasov gas onto black holes in bumblebee gravity \citep{Cai2022}. There are also numerous studies examining accretion onto Kerr black holes. Ref. \citep{Rioseco2018} illustrated that a collisionless, relativistic kinetic gas configuration propagating in the equatorial plane of a Kerr black hole would ultimately settle down to a stationary, axisymmetric configuration surrounding the black hole. In the slowly rotating case, Ref. \citep{Andersson2018} proved the decay of bounded energy and integrated energy for massless Vlasov gas in Kerr spacetime. Subsequently, Cie\'{s}lik \textit{et. al.} conducted a comprehensive study of the accretion of Vlasov gas onto a Kerr black hole in the equatorial plane \citep{Cieslik2022}. Ref. \citep{Mach2023} discussed the equatorial accretion on a moving Kerr black hole, while recently, a full $3+1$-dimensional accretion theory of Vlaso gas onto a Kerr black hole was established \citep{Li2023}. Moreover, some studies have started using numerical simulations to model Vlasov gas accretion processes. Cie\'{s}lik, Mach, and Odrzywo{\l}ek performed Monte Carlo simulations of stationary planar accretion of Vlasov gas onto a Schwarzschild black hole in Refs. \citep{Cieslik2023,Cieslik2024}.

The process of a black hole accreting a red giant is a significant event in astronomical observations \citep{Pacucci2022,Kareem2023}. Within red giants, degenerate ions are abundant and play a crucial role in their properties \citep{Deheuvels2022}. When a black hole at the center accretes the red giant, the degenerate ions would be pulled into the black hole's event horizon. Although true astrophysical plasmas are nearly electrically neutral, the theoretical importance of studying the accretion of charged Fermions is evident. First, when charged particles fall into a black hole, the electric charge of the black hole would change. Charged ions are inevitably present in interstellar matter and will be accreted by the black hole. Therefore, the electric charge of astronomical black holes actually changes from moment to moment over time. Secondly, even in the neutral accretion process, plasma may be composed of positive and negative ion pairs. The point particles that compose neutral plasma may also follow the Fermi-Dirac distribution. The literature \citep{Cieslik2020} only considers the accretion of neutral particles onto a Reissner-Nordstr\"{o}m (RN) black hole and does not address the change in the black hole's charge $Q$. So far, no articles have considered this issue in the Rioseco and Sarbach accretion model. In this paper, we delve into the accretion of charged particles onto black holes. Unlike other articles studying the Rioseco and Sarbach model, we use the Boyer-Lindquist coordinate system and show that the accretion of Vlasov gas is a two-perfect-fluid model. In order to alter the charge of a black hole, the particle falling into the black hole must carry a charge, too. As charged particles fall into the black hole, both the black hole's mass and charge change, as does the charge-to-mass ratio of the black hole. However, as shown in Ref. \citep{Wald1974}, the particles that would violate the horizon will not be captured by the black hole.

This paper is organized as follows: In Sec.II, we review the trajectory of charged particles in the Reissner-Nordstr\"{o}m black hole. In Sec.III, we discuss the general Vlasov equation in the Boyer-Lindquist coordinates of the Reissner-Nordstr\"{o}m background. In Sec.IV, we choose the Fermi-Dirac statistic as the distribution function at infinity. In Sec.V, we derive the expressions of the particle current density and the stress energy momentum tensor, both for the absorption part and the scattering part. In Sec.VI, we compare these quantities with those of an isotropic perfect fluid and found that our model describes an anisotropic fluid with two perfect-fluid components. In Sec.VII, we present three accretion rates: the mass accretion rate, the charge accretion rate, and the energy accretion rate. In Sec.VIII, we calculate the asymptotic behavior of observable quantities using Taylor series.
In Sec.IX, we numerically calculate the corresponding quantities in a finite range. The last section is a detailed conclusion of this paper. In this paper, we use units $G=c=1$.

\section{The motion of charged particle in Reissner-Nordstr\"{o}m black hole: Review}
In this paper, we consider the background to be Reissner-Nordstr\"{o}m black holes, which have the form
\begin{align}
  ds^2&=-\left(1-\frac{2M}{r}+\frac{Q^2}{r^2}\right)dt^2+\left(1-\frac{2M}{r}+\frac{Q^2}{r^2}\right)^{-1}dr^2\nonumber\\
  &+r^2d\theta^2+r^2\sin^2\theta d\varphi^2,
\end{align}
where $M$ is the mass and $Q$ is the charge of black hole. The outer horizon is $r_{+}=M+\sqrt{M^2-Q^2}$. The electromagnetic potential has components $A_{\mu}=(-\frac{Q}{r},0,0,0)$. The charged test particle should no longer describe a geodesic in the Reissner-Nordstr\"{o}m metric \citep{Chandrasekhar1983}. The motion of charged particles traveling in the Reissner-Nordstr\"{o}m geometry is determined by the Lagrangian
\begin{align}
  \mathcal{L} & =\frac{1}{2}g_{\mu\nu}\frac{dx^{\mu}}{d\lambda}\frac{dx^{\nu}}{d\lambda} +\frac{q}{m}A_{\mu}\frac{dx^{\mu}}{d\lambda}\nonumber\\
   & =-\frac{1}{2}\left(1-\frac{2M}{r}+\frac{Q^2}{r^2}\right)\dot{t}^2+\frac{1}{2}\left(1-\frac{2M}{r}+\frac{Q^2}{r^2}\right)^{-1}\dot{r}^2\nonumber\\
   &+\frac{1}{2}r^2\dot{\theta}^2+
   \frac{1}{2}r^2\sin^2\theta\dot{\varphi}^2-q\frac{Q}{r}\dot{t},
\end{align}
where $\lambda$ is the affine parameter and the dot denotes the differential with respect to $\lambda$. $m$ is the mass and $q$ is the charge of particle.

The canonical momentums $p_{\mu}$ are defined by
\begin{align}
p_t&=-\frac{\partial  \mathcal{L}}{\partial \dot{t}}=\left(1-\frac{2M}{r}+\frac{Q^2}{r^2}\right)\dot{t}+q\frac{Q}{r},\\
p_r&=\frac{\partial  \mathcal{L}}{\partial \dot{r}}=\left(1-\frac{2M}{r}+\frac{Q^2}{r^2}\right)^{-1}\dot{r},\\
p_{\theta}&=\frac{\partial  \mathcal{L}}{\partial \dot{\theta}}=r^2\dot{\theta},\\
p_{\varphi}&=\frac{\partial  \mathcal{L}}{\partial \dot{\varphi}}=r^2\sin^2\theta\dot{\varphi}.
\end{align}
Through the Legendre transformation, one can obtain the Hamiltonian
\begin{align}
\mathcal{H}&=p_{\mu}\dot{x}^{\mu}-\mathcal{L}\nonumber\\
&=\frac{1}{2}g_{\mu\nu}\frac{dx^{\mu}}{d\lambda}\frac{dx^{\nu}}{d\lambda}.
\end{align}
The equations of motion are given by
\begin{equation}\label{eom}
\frac{dx^{\mu}}{d\lambda} =\frac{\partial \mathcal{H}}{\partial p_{\mu}},\quad \frac{dp_{\mu}}{d\lambda} =-\frac{\partial \mathcal{H}}{\partial x^{\mu}}.
\end{equation}
Since the Hamiltonian does not contain $t$ and $\varphi$, the energy $E=-p_t$ and the angular momentum $l_z=p_{\varphi}$ in the $z$-direction are constants, which are
\begin{equation}
\dot{p}_t=-\frac{\partial \mathcal{H}}{\partial t}=0,\quad \dot{p}_{\varphi}=\frac{\partial \mathcal{H}}{\partial \varphi}=0.
\end{equation}
The Hamiltonian is conserved, defining another constant: the rest mass $m$
\begin{equation}
m^2=-2\mathcal{H}.
\end{equation}
The last constant in the spherical symmetric spacetime is the angular momentum $l^2$. To sum up, all four constants in this motion are
\begin{align}
m^2=&-2\mathcal{H}=-g_{\mu\nu}\frac{dx^{\mu}}{d\lambda}\frac{dx^{\nu}}{d\lambda},\\
E=&-\left(1-\frac{2M}{r}+\frac{Q^2}{r^2}\right)\dot{t}-q\frac{Q}{r},\\
l^2=&r^4\left((\frac{d\theta}{d\lambda})^2+\sin^2\theta(\frac{d\varphi}{d\lambda})^2\right),\\
l_z=&r^2\sin^2\theta\frac{d\varphi}{d\lambda}.
\end{align}
Using constants, the equations of motion (\ref{eom}) can be written as the decoupled, first-order equations
\begin{align}
\dot{t}&=-\left(1-\frac{2M}{r}+\frac{Q^2}{r^2}\right)^{-1}\left(E+q\frac{Q}{r}\right),\\
\dot{\varphi}&=\frac{l_z}{r^2\sin^2\theta},\\
\dot{r}^2&=-\left(1-\frac{2M}{r}+\frac{Q^2}{r^2}\right)(m^2+\frac{l^2}{r^2})+(E+q\frac{Q}{r})^2,\\
\dot{\theta}^2&=\frac{1}{r^4}\left(l^2-\sin^{-2}\theta l_z^2 \right).
\end{align}
For convenience to write, we define
\begin{align}
R&\equiv r^4 \dot{r}^2=(E^2-m^2)r^4+2(m^2M+EqQ)r^3\nonumber\\
&+((q^2-m^2)Q^2-l^2)r^2+2l^2Mr-l^2Q^2,\label{Rfunc}
\end{align}
which can be understood as the equivalent kinetic energy in the radial direction.

When the charged particles incident from infinity towards the Reissner-Nordstr\"{o}m black hole, the motions are bounded into a plane since the angular momentum $l^2$ is conserved. Similar to the geodesic motion in the Schwarzschild spacetime, we only consider the plane motion in $\theta=\frac{\pi}{2}$. At infinity, the equivalent kinetic energy $R>0$ indicates $E>m$. The equivalent kinetic energy $R$ also varied with respect to $r$. Since the kinetic energy can not be negative, particles only appeared in the range $R\geq0$. Let's define $r_0$ as the biggest real root of equation $R(r_0)=0$. A test particle can travel to $r=r_0$. There exist three cases to discuss, also see \citep{Li2023}.
\begin{enumerate}
  \item $r_0<r_+$: The test particle falls into the black hole.
  \item $r_0>r_+$, and $R<0$ for $r<r_0$: The test particle cannot enter the range $r<r_0$. It will be scattered by the black hole and travel to infinity. In this case, the position $r=r_0$ is the closest location, also known as the perihelion $r_p=r_0$, when particle traveling through the black hole.
  \item $r_0>r_+$, and $R>0$ for $r<r_0$: The test particle cannot travel to $r<r_0$ but keep getting close to $r=r_0$ and never reach there. The position $r=r_0$ stands for an unstable orbit $r_c=r_0$.
\end{enumerate}
Case 3 is a critical situation lying between case 1 and case 2. Since $R(r)>0$ is true for both $r>r_0$ and $r<r_0$, $R(r_0)=0$ is the minimum of function $R(r)$. Then, case 3 satisfies
\begin{align}
  R(r_c,l_c) &=0,\label{eq1}  \\
 \partial_{r}R(r_c,l_c) & =0.\label{eq2}
\end{align}
The above equations (\ref{eq1}) and (\ref{eq2}) are the key equations to classify the trajectories of charged particles.
\section{Action-angle variables and the Vlasov equation}
Let $\gamma$ be an orbit of charged particles traveling in the Reissner-Nordstr\"{o}m background. The abbreviated action is expressed as
\begin{equation}\label{aaction}
S=-Et+l_z\varphi+\int_{\gamma}p_rdr+\int_{\gamma}p_{\theta}d\theta,
\end{equation}
where the integral is along the orbit $\gamma$. Since $E$ contains $Q$, the electromagnetic interaction is already included in the abbreviated action.

Introducing the new momenta
\begin{align}
P_0&=m=\sqrt{-2\mathcal{H}},\\
P_1&=E=-p_t,\\
P_2&=l_z=p_{\varphi},\\
P_3&=l=\sqrt{p_{\theta}^2+\frac{p_{\varphi}^2}{\sin^2\theta}},
\end{align}
we can transform the canonical coordinates $(x^{\mu},p_{\mu})$ to the new canonical coordinates $(Q^{\mu},P_{\mu})$. The new conjugate coordinates are defined as
\begin{align}
Q^0&=\frac{\partial S}{\partial m}=-\int_{\gamma}\frac{r^2}{\sqrt{R}}\left(1-\frac{2M}{r}+\frac{Q^2}{r^2}\right)^{-1}\nonumber\\
&\cdot\left(m\left(1-\frac{2M}{r}+\frac{Q^2}{r^2}\right)-\frac{Q}{r}\frac{q}{m}\left(E+\frac{qQ}{r}\right)\right)dr,\\
Q^1&=\frac{\partial S}{\partial E}=-t+\int_{\gamma}\frac{r^2}{\sqrt{R}}\left(1-\frac{2M}{r}+\frac{Q^2}{r^2}\right)^{-1}\nonumber\\
&\cdot\left(E+\frac{qQ}{r}\right)dr,\\
Q^2&=\frac{\partial S}{\partial l_z}=\varphi-l_z\int_{\gamma}\frac{\sin^{-2}\theta }{\sqrt{l^2-\sin^{-2}\theta l_z^2}}d\theta,\\
Q^3&=\frac{\partial S}{\partial l}=-l\int_{\gamma}\frac{1}{\sqrt{R}}dr+l\int_{\gamma}\frac{d\theta}{\sqrt{l^2-\sin^{-2}\theta l_z^2}}.
\end{align}
The new canonical coordinates $(Q^{\mu},P_{\mu})$ are also known as the action-angle variables.

Assume that there are a large number of charged particles, and each of them travels along the orbit described by the abbreviated action (\ref{aaction}). At the same time, their behavior also can be treated as an ensemble, which is described by the distribution function $f(x^{\mu},p_{\nu})$. Furthermore, we consider that each particle interacts with others only through the gravitational and electromagnetic fields of the background. Thus, the distribution function $f(x^{\mu},p_{\nu})$ satisfied the Vlasov equation
\begin{equation}\label{Boltzmann}
  \dot{f}\equiv \{f,\mathcal{H}\}=0,
\end{equation}
where the bracket $\{.,.\}$ stands for the Poisson bracket.

The transformations $(x^{\mu},p_{\mu})\rightarrow (Q^{\mu},P_{\mu})$ keep the Poisson bracket (\ref{Boltzmann}) covariant
\begin{equation}
  \frac{\partial H}{\partial p_{\mu}}\frac{\partial}{\partial x^{\mu}}-\frac{\partial H}{\partial x^{\mu}}\frac{\partial}{\partial p_{\mu}}
  =  \frac{\partial H}{\partial P_{\mu}}\frac{\partial}{\partial Q^{\mu}}-\frac{\partial H}{\partial Q^{\mu}}\frac{\partial}{\partial P_{\mu}}.
\end{equation}
In the action-angle variables $(Q^{\mu},P_{\mu})$, the Hamiltonian becomes $H=\frac{P_0^2}{2}$. Therefore, the Vlasov equation (\ref{Boltzmann}) is simplified to
\begin{equation}
  \frac{\partial}{\partial Q^0}f=0,
\end{equation}
which is to say the distribution function $f$ does not involve $Q^0$. Using the symmetries, one can constrain the distribution function $f(x^{\mu},p_{\mu})$ to be a more simple form. This is first done by Ref. \citep{Rioseco2017-1}, which shows for the stationary spherical symmetric flow, the distribution function $f$ is independent of $Q^1,Q^2,Q^3$ and $P_2$. That is to say, the solution of the Vlasov equation in the stationary spherical symmetric case has a general form
\begin{equation}\label{disf}
f(x^{\mu},p_{\nu})=\mathcal{F}(P_0,P_1,P_3).
\end{equation}

\section{degenerate relativistic Fermi gas}
Astronomical observations often show black holes gradually pulling in massive objects, many of which are made of Fermions. To paint a more realistic picture of how black holes accrete matter, we examine a degenerate fluid composed of a large number of Fermions. At infinity, the Reissner-Nordstr\"{o}m metric is asymptotically flat and we assume that the Fermi gas is in thermal equilibrium. Due to the black hole's gravitational pull, the Fermi gas should be limited to a finite range along its trajectory. Fermions adhere to Pauli's exclusion principle, meaning that each quantum state of an electron is allowed to be occupied by only one particle. The statistical law that describes their behavior at infinity is known as Fermi-Dirac statistics, which is given by the expression \citep{Rezzolla2013}
\begin{equation}\label{fer}
  f_{\infty}=A\frac{1}{\exp[(E-\mu)/k_BT]+1},
\end{equation}
where $A$ is the normalization factor. $k_B$ is the Boltzmann constant, $T$ is the temperature and $\mu$ is the relativistic chemical potential. The Fermi-Dirac distribution (\ref{fer}) usually requires a high temperature and in particular a heat bath at equilibrium. Inside the white dwarf or neutron star, an extremely high pressure and temperature can provide such a heat bath. In order to satisfy this condition, we consider the black hole accretes the white dwarf or neutron star at a finity distance. Imagine that this compact object was captured from infinity by the gravitational pull of the black hole. Then, the compact object of any finite distance can be pushed back to infinity by the gravitational capture trajectory. Thus, such a heat bath could exist at infinity. \footnote{In the finity distance, it doesn't important whether there's a heat bath or not. This is because at a finite distance, point particles of the Rioseco and Sarbach model are not in equilibrium. They simply move along the geodesics of the black hole.} There could be some apprehension. The above distribution function (\ref{fer}) is often used to describe electrically neutral Fermi systems. If particles have a charge $q$, electromagnetic interactions between point particles may disrupt the distribution. However, we emphasize that the Vlasov equation (\ref{Boltzmann}) is only valid when ignoring the electromagnetic interactions between particles. If the particle's electromagnetic interaction is taken into account, the Poisson bracket (\ref{Boltzmann}) would no longer be zero, and the distribution function (\ref{fer}) would change. Therefore, the selection of distribution function (\ref{fer}) needs to consider the conditions of Vlasov equation. The above distribution function is valid in at least two cases. The first case is an electrically neutral Fermion gas, such as a neutron gas. The second case is the zero-order effect of the electron gas. Since the interaction term is related to the charge $q$, it is sufficient to consider the zero-order effects when $q$ is small. 
 
Similar to Ref. \citep{Rioseco2017-1}, the distribution function throughout all spacetime is giving by
\begin{equation}\label{Fermidis}
  f=\delta(P_0-m_0)f_{\infty}=A\delta(P_0-m_0)\frac{1}{\exp[(P_1-\mu)/k_BT]+1},
\end{equation}
where $m_0$ is the mass of Fermions. The above distribution function $f$ is a specific solution of Vlasov equation and obeys Eq.(\ref{disf}). It is valid everywhere, not only at infinity. However, in a finite radius, $T$ is only a parameter and does not stand for temperature.

\section{Observable quantities}
The observable quantities we care about are the particle current density $J_{\mu}$ and the stress energy-momentum tensor $T_{\mu\nu}$. They are given by a suitable integral over momenta
\begin{align}
J_{\mu}&=\int p_{\mu}f(x,p)\mathrm{dvol}_x(p),\label{Ober1}\\
T_{\mu\nu}&=\int p_{\mu}p_{\nu}f(x,p)\mathrm{dvol}_x(p),\label{Ober2}
\end{align}
where the volume element is given by
\begin{equation}
\mathrm{dvol}_x(p)=\sqrt{-\det[g^{\mu\nu}]}dp_0dp_1dp_2dp_3.
\end{equation}
Using Eq. (\ref{Boltzmann}), one can easily show that the particle current density $J_{\mu}$ and the energy-momentum tensor $T_{\mu\nu}$ satisfy the conservation laws
\begin{align}
  \nabla_{\mu}J^{\mu} &=0,  \\
   \nabla_{\mu}T^{\mu\nu}& =0.
\end{align}

Notice that the observable quantities are actually calculated in the action-angle variables $(P_0,P_1,P_2,P_3)$. We then transform the momentum variables $(p_t,p_r,p_{\theta},p_{\varphi})$ to the variables $(m^2,E,l_z,l^2)$, which are expressed as follow
\begin{align}
m^2 &= -g^{00}(p_t-\frac{q}{m}\frac{Q}{r})^2-g^{11}p_r^2-g^{22}p_{\theta}^2-g^{33}p_{\varphi}^2,\\
E &= -p_t,\\
l_z &= p_{\varphi},\\
l^2 &= p_{\theta}^2+\sin^{-2}\theta p_{\varphi}^2.
\end{align}
The Jacobian determinant is defined by
\begin{equation}
J=\frac{\partial(m^2,E,l_z,l^2)}{\partial(p_t,p_r,p_{\theta},p_{\varphi})}=\frac{4\sqrt{R}}{r^2}\sqrt{l^2-\sin^{-2}\theta l_z^2}.
\end{equation}
Therefore, the volume element can be reexpressed as
\begin{equation}
\mathrm{dvol}_x(p)=\frac{1}{4\sqrt{R}}\frac{1}{\sqrt{l^2\sin^{2}\theta- l_z^2}}dm^2dEdl_zdl^2.
\end{equation}
In order to keep the function inside the square root non-negative, the integral range of $l_z$ is $l_z\in[-l\sin\theta,l\sin\theta]$. We further define
\begin{equation}
l_z=l\sin\theta\sin\sigma,
\end{equation}
and transform the integral of $l_z$ to the integral of $\sigma$. It is easy to obtain
\begin{equation}
\mathrm{dvol}_x(p)=\frac{1}{4\sqrt{R}}dm^2dEd\sigma dl^2,
\end{equation}
and the integral range of $\sigma$ is $\sigma\in[0,2\pi]$.

According to the classification of trajectories, the observable quantities $J_{\mu}$ and $T_{\mu\nu}$ can be divided into three parts: the absorbed part, the scattered part and the critical part, which are
\begin{align}
  J_{\mu} &=J^{abs}_{\mu}+J^{scat}_{\mu}+J^{cri}_{\mu},  \\
  T_{\mu\nu} &=T^{abs}_{\mu\nu}+T^{scat}_{\mu\nu}+T^{cri}_{\mu\nu}.
\end{align}
The integral interval of the parameter $(m,E,\sigma,l)$ will determine which part the quantities $J_{\mu}$ and $T_{\mu\nu}$ belong to. Specially, the critical part happens to be $l=l_{c}$. Thus, the integration  $\int_{l_{c}}^{l_{c}}f(l)dl$ vanishes and the critical part does not have a substantial contribution.

\subsection{Absorption part}
For the absorption part, there is no perihelion on the orbit. Solving the simultaneous Eq. (\ref{eq1}) and (\ref{eq2}), one obtain
\begin{equation}
r=r_c(E),\quad
l=l_c(E),
\end{equation}
where $r_c(E)$ is the unstable orbit. Notice that the expression (\ref{Rfunc}) of $R$ does not contained $l_z$, the solutions $(r_c(E),l_c(E))$ are independent of $\sigma$. $l_c(E)$ determined whether there is a perihelion. For trajectories with parameter $l>l_c$, the perihelion exists and the charged particles should be scattered to infinity; otherwise, the charged particles would be captured by the black hole. Thus, for the absorption part, the integral interval is $l\in[0,l_c)$.

Plugging the distribution function (\ref{Fermidis}) into the observable quantities (\ref{Ober1}), we obtain the absorption part of $J_{\mu}$ as follows
\begin{align}
J^{abs}_{t}&=-2\pi Am_0\int_{m_0}^{\infty}dE\int_0^{l_c}dl\frac{El}{\sqrt{R_0}}F,\\
J^{abs}_{r}&=2\pi Am_0\int_{m_0}^{\infty}dE\int_0^{l_c}dl\frac{l}{(r^2-2Mr+Q^2)}F,
\end{align}
where  $F=\frac{1}{(\exp[(E-\mu)\beta]+1)},\beta=\frac{1}{k_bT}$ and $R_0$ stands for the function $R$ in the case of $m=m_0$. Since $p_{\theta}=l\sin\sigma$ and $p_{\varphi}=l\sin\theta\sin \sigma$, integrating $\sigma$ in a period, one can easily show $J^{abs}_{\theta}$ and $J^{abs}_{\varphi}$ vanish.

For the same reason, one can also show that $T_{t\theta}=T_{t\varphi}=T_{r\theta}=T_{r\varphi}=T_{\theta\varphi}=0$. The non-zero components of $T_{\mu\nu}$ are expressed as
\begin{align}
T^{abs}_{tt}&=2\pi Am_0\int_{m_0}^{\infty}dE\int_0^{l_c}dl\frac{E^2l}{\sqrt{R_0}}F,\\
T^{abs}_{tr}&=-2\pi Am_0\int_{m_0}^{\infty}dE\int_0^{l_c}dl\frac{El}{(r^2-2Mr+Q^2)}F ,\\
T^{abs}_{rr}&=2\pi Am_0\int_{m_0}^{\infty}dE\int_0^{l_c}dl\frac{\sqrt{R_0}l}{(r^2-2Mr+Q^2)^2}F,\\
T^{abs}_{\theta\theta}&=\frac{ T^{abs}_{\varphi\varphi}}{\sin^2\theta}=\pi Am_0\int_{m_0}^{\infty}dE\int_0^{l_c}dl\frac{l^3}{\sqrt{R_0}}F.
\end{align}

\subsection{Scattering part}
For the scattering part, the condition for the existence of perihelion requirements $l>l_c$. At the same time, the signature of $R_0$ must be positive. Solving $R_0=0$, one obtains the maximum of $l_{max}(E)$, which is
\begin{equation}
l_{max}=r\sqrt{\frac{r^2}{r^2-2Mr+Q^2}\left(E+\frac{qQ}{r}\right)^2-m_0^2}.
\end{equation}
Thus, the integral interval of $l$ is $l\in (l_c,l_{max}]$.

The integral of $E$ is more complicated. The main reason is that the closer a particle gets to the black hole, the more energy it takes to escape. We rewrite $R_0$ into
\begin{equation}
R_0=r^4\left((E+q\frac{Q}{r})^2-\left(1-\frac{2M}{r}+\frac{Q^2}{r^2}\right)(m_0^2+\frac{l^2}{r^2}) \right),
\end{equation}
and define the potential
\begin{equation}
U_l(r)=\left(1-\frac{2M}{r}+\frac{Q^2}{r^2}\right)(m_0^2+\frac{l^2}{r^2}).
\end{equation}
Since $R_0\geq 0$, the energy $E$ required for a particle to escape the black hole satisfy $E\geq\sqrt{U_{l}}-\frac{qQ}{r}$. We define
\begin{align}
\Psi_{l}(r)&=\sqrt{U_{l}}-\frac{qQ}{r}.
\end{align}
In the maximum of $\Psi_{l}(r)$, i.e. $\frac{d \Psi_{l}(r)}{d r}=0$, there is
\begin{align}
l^2_{\pm}&=\frac{1}{2Y^2}\bigg(2Y(\Delta-Y)m_0^2r^2+q^2Q^2r^2\Delta\nonumber\\
& \pm qQ r^2\Delta\sqrt{4Ym_0^2+q^2Q^2} \bigg),
\end{align}
where
\begin{align}
\Delta(r)&=r^2-2M r+Q^2=(r-r_{+})(r-r_{-}),\\
Y(r)&=r^2-3M r+2Q^2=(r-r_{ph+})(r-r_{ph-}).
\end{align}
The expressions $r_{\pm}=M\pm\sqrt{M^2-Q^2}$ are the outer or inner horizon and $r_{ph\pm}=\frac{1}{2}(3M\pm\sqrt{9M^2-4Q^2})$ are the radii of photo orbit. In the special case of $q=0$, we reproduce the result in ref. \citep{Cieslik2020}
\begin{equation}
l^2_{+}=l^2_{-}=\frac{(\Delta-Y)}{Y}m_0^2r^2.
\end{equation}
Inserting $l^2_{+}$ into $\Psi_{l}$, one have
\begin{equation}
\Psi_{l_+}=\frac{m_0\Delta}{r\sqrt{Y}}\sqrt{1+\frac{1}{2Y}\frac{qQ}{m_0}\left(\frac{qQ}{m_0}-\sqrt{4Y+\frac{q^2Q^2}{m_0^2}}\right)}-\frac{qQ}{r}.
\end{equation}
As $r$ approaching to outer phone sphere $r\rightarrow r_{ph+}$, the function $\Psi_{l_+}\rightarrow\infty$. That is to say, there is no particle that can escape the black hole. On the other hand, for particles that can incident from infinity, their energy satisfies $E\geq m_0$. Or equivalent to say, the minimal energy always satisfies $E_{min}\geq m_0$. Therefore, the radial coordinate is divided into three segments
\begin{equation}
E_{min}=\left\{
          \begin{array}{ll}
            \infty, &    \hbox{$r<r_{ph+}$;} \\
           \Psi_{l_+} , & \hbox{$r_1>r>r_{ph+}$;} \\
            m_0, & \hbox{$r>r_1$.}
          \end{array}
        \right.
\end{equation}
where $r_1$ is the root of $\Psi_{l_+}(r_1)=m_0$. At last, we obtain that the integral interval of $E$ is $E\in[E_{min},\infty)$.

Notice that the trajectory of the scattering part contains both $+\sqrt{R_0}$ and $-\sqrt{R_0}$. Therefore, we expressed the non-zero observable quantities as follows
\begin{align}
J^{scat}_{t}&=-2\pi Am_0\sum_{\pm}\int_{E_{min}}^{\infty}dE\int_{l_c}^{l_{max}}dl\frac{El}{\sqrt{R_0}}F,\\
J^{scat}_{r}&=2\pi Am_0\sum_{\pm}\int_{E_{min}}^{\infty}dE\int_{l_c}^{l_{max}}dl\frac{l}{(r^2-2Mr+Q^2)}F,
\end{align}
\begin{align}
T^{scat}_{tt}&=2\pi Am_0\sum_{\pm}\int_{E_{min}}^{\infty}dE\int_{l_c}^{l_{max}}dl\frac{E^2l}{\sqrt{R_0}}F,\\
T^{scat}_{tr}&=-2\pi Am_0\sum_{\pm}\int_{E_{min}}^{\infty}dE\int_{l_c}^{l_{max}}dl\frac{El}{(r^2-2Mr+Q^2)}F,\\
T^{scat}_{rr}&=2\pi Am_0\sum_{\pm}\int_{E_{min}}^{\infty}dE\int_{l_c}^{l_{max}}dl\frac{\sqrt{R_0}l}{(r^2-2Mr+Q^2)^2}F,\\
T^{scat}_{\theta\theta}&=\frac{ T^{scat}_{\varphi\varphi}}{\sin^2\theta}=\pi Am_0\sum_{\pm}\int_{E_{min}}^{\infty}dE\int_{l_c}^{l_{max}}dl\frac{l^3}{\sqrt{R_0}}F.
\end{align}

\section{Comparison with the perfect fluid case}
In this section, we make a comparison between the observable quantities of Vlasov gas with those of an isotropic perfect fluid. In the perfect fluid model, the particle current density and the stress energy-momentum tensor are given by
\begin{equation}
J^{\mu}=nu^{\mu},\quad T^{\mu}{}_{\nu}=\rho u^{\mu}u_{\nu}+p(u^{\mu}u_{\nu}+\delta^{\mu}_{\nu}),
\end{equation}
where $n$ is particle density, $u^{\mu}$ is four-velocity, $\rho$ is the energy density and $p$ is the pressure. As is shown in Ref. \citep{Cieslik2020}, $u^{\mu}$ is an eigenvector of $T^{\mu}{}_{\nu}$ corresponding to the eigenvalue $-\rho$. For any non-zero $k^{\mu}$ orthogonal to $u_{\mu}$ is also an eigenvector corresponding to the eigenvalue $p$. In the perfect fluid model, the pressure $p$ is a threefold degenerate eigenvalue of $T^{\mu}{}_{\nu}$.

In the Vlasov case, things would be very different. Since $T_{\theta\varphi}=0$ and $T_{\varphi\varphi}=\sin^2\theta T_{\theta\theta}$, there existed twofold eigenvalue $p_{\mathrm{tan}}=T^{\theta}{}_{\theta}=T^{\varphi}{}_{\varphi}$ for $T^{\mu}{}_{\nu}$, corresponding to the eigenvector $(0,0,1,0)$ and $(0,0,0,1)$ respectively. In order to calculate the other two eigenvectors, we consider the matrix $\left(
  \begin{array}{cc}
    T^t{}_t & T^t{}_r \\
    T^r{}_t & T^r{}_r \\
  \end{array}
\right)$. Calculating the secular equation, one obtains two eigenvalues
\begin{equation}
\lambda_{\pm}=\frac{T^t{}_t+T^r{}_r\pm\sqrt{(T^t{}_t-T^r{}_r)^2+4 T^t{}_r T^r{}_t}}{2},
\end{equation}
which corresponding to the energy density $-\rho$ and the radial pressure $p_{\mathrm{rad}}$. Thus, the general form of energy-momentum tensor is expressed as
\begin{equation}\label{Twoper}
  T^{\mu}{}_{\nu}=\left(
               \begin{array}{cccc}
                 -\rho & 0 & 0 & 0 \\
                 0 & p_{\mathrm{rad}} & 0 & 0 \\
                 0 & 0 & p_{\mathrm{tan}} & 0 \\
                 0 & 0 & 0 & p_{\mathrm{tan}} \\
               \end{array}
             \right).
\end{equation}
In the case of $p_{\mathrm{rad}}\neq p_{\mathrm{tan}}$, as is pointed in Ref. \citep{Letelier1980}, this is an anisotropic fluid with two-perfect-fluid components. One can construct the above energy-momentum tensor (\ref{Twoper}) by combining two perfect fluids, one perfect and one null fluid, or two null fluids. In the case of Vlasov gas, the energy-momentum tensor can be decoupled into one perfect fluid corresponding to the Fermi gas and another null fluid. It is the null fluid that distinguishes the Vlasov model from the perfect fluid model. Latter, we will calculate the eigenvalues $\rho, p_{\mathrm{rad}}, p_{\mathrm{tan}}$ as the function of $r$.

\section{Accretion rates}
For the spherical symmetric flow, there are two accretion rates: one is the particle accretion rate $\dot{n}$, meaning number of particles that cross the horizon per unit time; the other is the energy accretion rate $\dot{E}$, meaning total energy crossing the horizon per unit time. The particle number is defined by 
\begin{equation}\label{pnumb}
n=\sqrt{-g_{\mu\nu}J^{\mu}J^{\nu}}.
\end{equation}
Then, the particle accretion rate $\dot{n}$ is also defined by the particle current density $J^{\mu}$. Notice that $\nabla_{\mu}J^{\mu}=0$ becomes
\begin{equation}
\frac{\partial}{\partial r}(r^2\sin\theta J^{r})=0.
\end{equation}
Let $V$ be the region of a black hole and $S$ is the surface of the outer horizon. The vector $\hat{n}=(1,0,0)$ is the unit normal vector of $S$ directed outside. Integrating the above equation and using the Stokes' theorem, the particle accretion rate $\dot{n}$ is defined by
\begin{align}
\dot{n}&=-\int_{V}(r^2\sin\theta J^{r})_{,r}dV\nonumber\\
&=-\int_{S} J^{r}(r_+)\hat{n}_{r}r_{+}^2\sin\theta d\theta d\varphi\nonumber\\
&=-4\pi r_{+}^2 J^{r}(r_{+}).
\end{align}

When an electron falls into the black hole, its mass $m_0$ and charge $q$ are absorbed by the black hole, so that the black hole's mass increases by $m_0$ and charge increases by $q$. Therefore, the growth of mass and charge for the black hole is caused by the particle number which falls into the black hole. We then define the
mass accretion rates $\dot{M}$ and the charge accretion rates $\dot{Q}$ as
\begin{align}
\dot{M}&=m_0\dot{n}=-4\pi r_{+}^2 m_0 J^{r}(r_{+}),\\
\dot{Q}&=q\dot{n}=-4\pi r_{+}^2 q J^{r}(r_{+}).
\end{align}

Similar to $J^{\mu}$, the stress energy momentum tensor $T_{\mu\nu}$ are also divergence free. Then, one can define the energy accretion rate
\begin{equation}
\dot{E}=-4\pi r_{+}^2 T^{r}{}_{t}(r_{+}).
\end{equation}
\section{Asymptotic behavior}
In this section, we discussed the asymptotic behavior of $J^{\mu}$ and $T_{\mu\nu}$ by using the Taylor series.
At large distances, the absorption part becomes zero, and only contributions of the scattering part remain. We rewrite $R_0$ to
\begin{equation}
R_0(r)=X(r)+\Delta(r) l^2,
\end{equation}
where
\begin{align}
X(r)&=(E^2-m_0^2)r^4+2(Mm_0^2+EqQ)r^3+\left(q^2-m_0^2\right)Q^2r^2.
\end{align}
The following integration of $l$ is obtained
\begin{equation}
\int_{l_c}^{l_{max}}\frac{l}{\sqrt{R_0}}dl=\int_{l_c}^{l_{max}}\frac{l}{\sqrt{X-\Delta l^2}}dl=\frac{\sqrt{X-\Delta l_c^2}}{\Delta},
\end{equation}
where we have used the identity $X+\Delta l_{max}^2=0$. Notice that $l_{max}$ can be expand as $\frac{1}{r}$
\begin{equation}
 l_{max}^2=r^2\left(E^2-m_0^2+2(E^2M+EqQ)\frac{1}{r} +o(\frac{1}{r})\right),
\end{equation}
But $l_c$ does not depend on $r$. Then, one obtains
\begin{equation}
\int_{l_c}^{l_{max}}\frac{l}{\sqrt{R_0}}dl=\sqrt{E^2-m_0^2}\left(1+\frac{2E^2M+EqQ-m_0^2M}{(E^2-m_0^2)}\frac{1}{r}\right).
\end{equation}
Therefore, using the Taylor expansion of $\frac{1}{r}$, we obtain
\begin{align}
J_t(r\rightarrow \infty)&=-4\pi Am_0\int_{E_{min}}^{\infty}dE\int_{l_c}^{l_{max}}dl\frac{El}{\sqrt{R_0}}F\nonumber\\
&=-4\pi Am_0\int_{m_0}^{\infty}\frac{\sqrt{E^2-m_0^2}E}{\exp[(E-\mu)\beta]+1}\bigg(1\nonumber\\
&+\frac{2E^2M+EqQ-m_0^2M}{E^2-m_0^2}\frac{1}{r}\bigg)dE,\label{Jinfty}
\end{align}
where we only list the first two terms.

The function $\frac{1}{\sqrt{R_0}}$ can also be expanded in terms of $\frac{1}{r}$
\begin{equation}
\frac{1}{\sqrt{R_0}}=\frac{1}{\sqrt{E^2-m_0^2}}\left(\frac{1}{r^2}-\frac{q Q E+m_0^2M}{E^2-m_0^2}\frac{1}{r^3}+o(\frac{1}{r^3})\right).
\end{equation}
Through a similar calculation, we obtain
\begin{align}
J_{r}(r\rightarrow \infty)&=2\pi Am_0\int_{m_0}^{\infty}\frac{E^2-m_0^2}{\exp[(E-\mu)\beta]+1}\bigg(1\nonumber\\
&+\frac{2E^2M+EqQ-m_0^2M}{E^2-m_0^2}\frac{2}{r}\bigg)dE,\\
T_{tt}(r\rightarrow \infty)&=4\pi Am_0\int_{m_0}^{\infty}\frac{\sqrt{E^2-m_0^2}E^2}{\exp[(E-\mu)\beta]+1}\bigg(1\nonumber\\
&+\frac{2E^2M+EqQ-m_0^2M}{E^2-m_0^2}\frac{1}{r}\bigg)dE,\\
T_{tr}(r\rightarrow \infty)&=-2\pi Am_0\int_{m_0}^{\infty}\frac{(E^2-m_0^2)E}{\exp[(E-\mu)\beta]+1}\bigg(1\nonumber\\
&+\frac{2E^2M+EqQ-m_0^2M}{E^2-m_0^2}\frac{2}{r}\bigg)dE,\\
T_{rr}(r\rightarrow \infty)&=\frac{4}{3}\pi Am_0\int_{m_0}^{\infty}\frac{(E^2-m_0^2)^{\frac{3}{2}}}{\exp[(E-\mu)\beta]+1}\bigg(1\nonumber\\
&+\frac{2E^2M+EqQ-m_0^2M}{E^2-m_0^2}\frac{3}{r}\bigg)dE,\\
T_{\theta\theta}(r\rightarrow \infty)&=\frac{T_{\varphi\varphi}(r\rightarrow \infty)}{\sin^2\theta}\nonumber\\
&=\frac{4r^2}{3}\pi Am_0\int_{m_0}^{\infty}\frac{(E^2-m_0^2)^{\frac{3}{2}}}{\exp[(E-\mu)\beta]+1}\bigg(1\nonumber\\
&+\frac{4E^2M+3EqQ-m_0^2M}{E^2-m_0^2}\frac{1}{r}\bigg)dE.\label{Tinfty}
\end{align}
It is not difficult to find, the stress energy-momentum tensor $T_{\mu\nu}$ at infinity satisfy
\begin{equation}
T_{\infty}{}^{r}{}_{r}=T_{\infty}{}^{\theta}{}_{\theta}=T_{\infty}{}^{\varphi}{}_{\varphi}.
\end{equation}

For the classical particles which satisfied Maxwell-J\"{u}ttner distribution, the above quantities turn into
\begin{align}
J_{\infty t}&=-4\pi A m_0^4\frac{K_2(z)}{z},\\
J_{\infty r}&=4\pi A m_0^4\frac{1+z}{z^3}e^{-z},\\
T_{\infty}{}^{t}{}_{t}&=4\pi A m_0^5\left(\frac{K_1(z)}{z}+3\frac{K_2(z)}{z^2}\right),\\
T_{\infty}{}^{t}{}_{r}&=-T_{\infty}{}^{r}{}_{t}=4 \pi A m_0^5 \frac{z^2+3z+3}{z^4}e^{-z},\\
T_{\infty}{}^{r}{}_{r}&=T_{\infty}{}^{\theta}{}_{\theta}=T_{\infty}{}^{\varphi}{}_{\varphi}=4\pi A m_0^5 \frac{K_2(z)}{z^2},
\end{align}
where $K_1(x)$ is the first Bessel function, $K_2(x)$ is the secondary Bessel function and $z=m_0\beta$. Since $T_{\infty}{}^{t}{}_{r}=-T_{\infty}{}^{r}{}_{t}$ is not vanish, the Vlasov gas is actually an anisotropic fluid with two perfect fluids even at infinity. The energy-momentum tensor for the Fermi gas is an isotropic perfect fluid, which has a form  
\begin{equation}
  T^{(fer)}_{\infty}{}^{\mu}{}_{\nu}=4\pi A m_0^5\left(
                                      \begin{array}{cccc}
                                        \frac{K_1(z)}{z}+3\frac{K_2(z)}{z^2} & 0 & 0 & 0 \\
                                        0 & \frac{K_2(z)}{z^2} & 0 & 0 \\
                                        0 & 0 & \frac{K_2(z)}{z^2} & 0 \\
                                        0 & 0 &0 & \frac{K_2(z)}{z^2} \\
                                      \end{array}
                                    \right).
\end{equation}
Another component is a null fluid and has the energy-momentum tensor 
\begin{equation}
 T^{(null)}_{\infty}{}_{\mu\nu}=4 \pi A m_0^5 \frac{z^2+3z+3}{z^4}e^{-z}(\delta^{0}_{\mu}\delta^{1}_{\nu}+\delta^{1}_{\mu}\delta^{0}_{\nu}).
\end{equation}
When the temperature approaching to zero $z\rightarrow \infty$, i.e. $T_{\infty}{}^{t}{}_{r}=-T_{\infty}{}^{r}{}_{t}\rightarrow 0$, the collisionless gas behaves an ideal fluid at infinity. Then, the particle number $n$ at large distance mainly contributed by $J_{t}$
\begin{equation}
n_{\infty}= 4\pi A m_0^4\frac{K_2(z)}{z}.
\end{equation}
For the same reason, the collisionless Fermi gas also can not be treated as the ideal fluid at infinity. The radial pressure $p_{\mathrm{rad}}$ is not equal to the tangential pressure $p_{\mathrm{tan}}$.

The above equations (\ref{Jinfty}) - (\ref{Tinfty}) also indicate that the physical results are independent of the parameters $q$ and $Q$ at the leader term of Taylor's expansion. However, the distribution function affects significantly. When performing the numerical results, this behavior is verified.

At the horizon, the scattering part is omitted and only the absorption part contributed. Eq.(\ref{eq1}) and (\ref{eq2}) only permit numerical solution for $l_c$. In the Schwarzschild limit $Q=0$, $l_c$ have the analytical expression, that is
\begin{equation}
l_c^2=\frac{M^2}{2}\frac{27E^4-36E^2m_0^2+8m_0^4+E(9E^2-8m_0^2)^{\frac{3}{2}}}{E^2-m_0^2}.
\end{equation}
Thus, for the Schwarzschild black hole to accrete the collisionless Fermi gas, there is
\begin{equation}
\frac{\dot{n}}{n_{\infty}}\bigg|_{Sch}=\pi\frac{\int_{m_0}^{\infty}\frac{l_c^2}{\exp[(E-\mu)\beta]+1}dE}{\int_{m_0}^{\infty}\frac{\sqrt{E^2-m_0^2}E}{\exp[(E-\mu)\beta]+1}dE}.
\end{equation}

\begin{figure}
\resizebox{0.45\textwidth}{!}{%
  \includegraphics{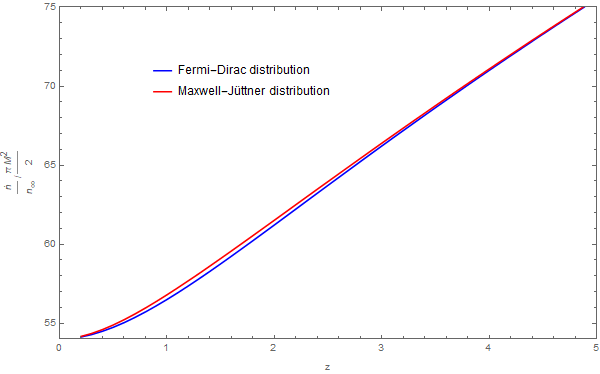}
}
\caption{The numerical results $\frac{\dot{n}}{n_{\infty}}\big|_{Sch}$ for the accretion onto a Schwarzschild black hole, where the unit is been chosen as $\frac{\pi M^2}{2}$.}
\label{Fig:NumAcre}
\end{figure}

We plot the numerical results of $\frac{\dot{n}}{n_{\infty}}\big|_{Sch}$ both for the Fermi gas and the classical gas in Fig.\ref{Fig:NumAcre}. In the low temperature limit $z\rightarrow \infty$, the number $\frac{\dot{n}}{n_{\infty}}\big|_{Sch}$ equal to the same value.

\section{Numerical results}
\subsection{particle current density and particle density}
\begin{figure}
\resizebox{0.45\textwidth}{!}{%
  \includegraphics{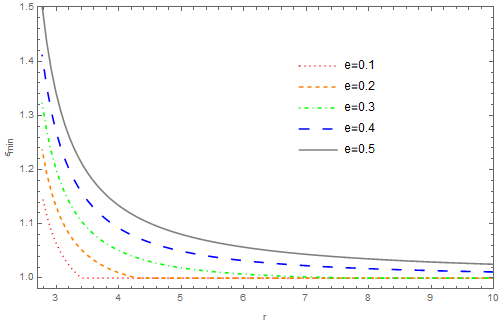}
}
\caption{$\varepsilon_{min}$ as the function of $r$. We choose $Q=0.9,\mu=0,z=1$.}
\label{Fig:Emin}
\end{figure}
The numerical calculation is performed by Wolfram Mathematica. We introduce the following dimensionless variables
\begin{equation}
\varepsilon=\frac{E}{m_0},\quad \xi=\frac{l}{m_0},\quad e=\frac{q}{m_0}.
\end{equation}
Without loss of generality, we choose $M=1$ and $e\leq 1, Q\leq1$. The function $\varepsilon_{min}(r)=\frac{E_{min}(r)}{m_0}$ is plotted in Fig.\ref{Fig:Emin}, which show that $\varepsilon_{min}$ grows as $e$ increasing. As a numerical calculation trick, we don't have to solve the equation $\Psi_{l_+}=m_0$ to derive $r_1$, but choose the maximum of $\max\{\frac{\Psi_{l_+}}{m_0},1\}$ to obtain $\varepsilon_{min}$.

\begin{figure}
\resizebox{0.45\textwidth}{!}{%
  \includegraphics{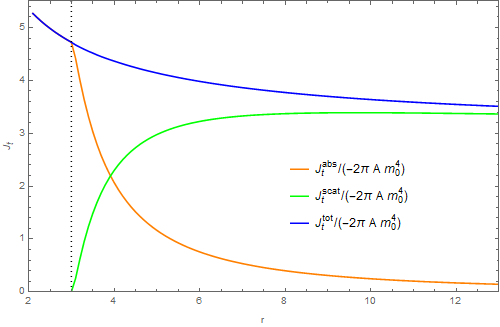}
}
\resizebox{0.45\textwidth}{!}{%
  \includegraphics{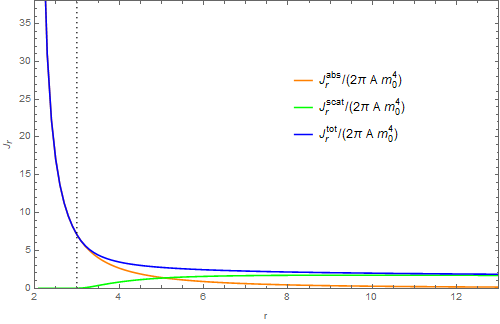}
}
\caption{Sample graphs of $J_t/(-2\pi A m_0^4)$ (up) and $J_r/(2\pi A m_0^4)$ (down). The parameters are chosen to be $Q=0.1,e=0.1,\mu=0,z=1$. The vertical dotted line represents the location of the photon sphere.}
\label{Fig:JtJr}
\end{figure}

\begin{figure}
\resizebox{0.45\textwidth}{!}{%
  \includegraphics{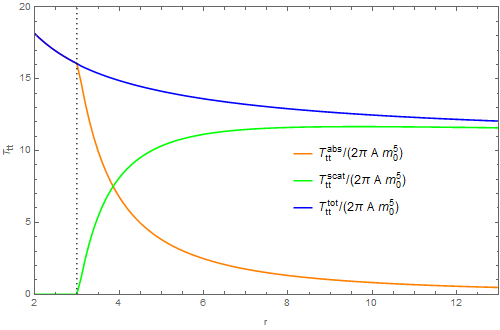}
}
\resizebox{0.45\textwidth}{!}{%
  \includegraphics{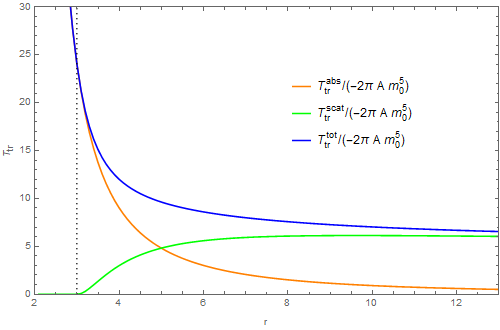}
}
\resizebox{0.45\textwidth}{!}{%
  \includegraphics{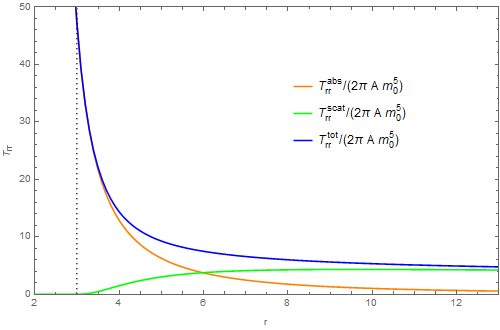}
}
\caption{Sample graphs of $T_{tt}/(2\pi A m_0^5)$ (up), $T_{tr}/(-2\pi A m_0^5)$ (middle) and $T_{rr}/(2\pi A m_0^5)$ (down). The parameters are chosen to be $Q=0.1,e=0.1,\mu=0,z=1$.}
\label{Fig:T}
\end{figure}

Using the function $\varepsilon_{min}$, we numerical calculate $J_{t}$ and $J_{r}$ as the function of $r$, see Fig.\ref{Fig:JtJr}. The absorption part and the scattering part are calculated separately. The scattering part vanishes when $r$ enters inside the photon sphere. The absorption portion is also discontinuous in the photon sphere. However, the total part is smooth in the photon sphere. Since we using the Boyer-Lindquist coordinate system, there is a singularity at the event horizon. The numerical calculation is performed outside the horizon. When $r\rightarrow r_{+}$, Fig.\ref{Fig:JtJr} shows that $J_{r}$ approaches infinity. However, as we will see later, this does not affect the calculation of the accretion rate. We also plot $T_{tt}$, $T_{tr}$ and $T_{rr}$ as the function of $r$ in Fig.\ref{Fig:T}. Their behavior is similar to $J_{\mu}$.

\begin{figure}
\resizebox{0.45\textwidth}{!}{%
  \includegraphics{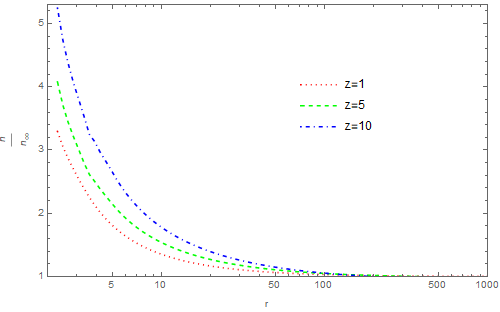}
}
\caption{Sample graphs of $\frac{n}{n_{\infty}}$. The parameters are chosen to be $Q=0.2,e=0.6,\mu=0,z=1$.}
\label{Fig:N}
\end{figure}

The particle number density $n$ is expressed as Eq.(\ref{pnumb}). Using the numerical results of $J_{\mu}$, we plot the particle number density $\frac{n}{n_{\infty}}$ as the function of $r$ in Fig.\ref{Fig:N}. As $z$ increases, the function $\frac{n}{n_{\infty}}$ grows near the horizon and still approaches 1 far away. A similar behavior is also shown by Ref. \citep{Rioseco2017-1,Cieslik2020}.

\subsection{the cosmic censorship and accretion rates }
\begin{figure}
\resizebox{0.45\textwidth}{!}{%
  \includegraphics{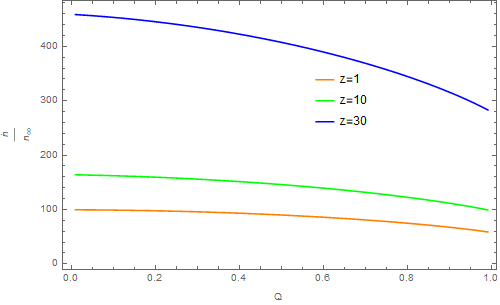}
}
\resizebox{0.45\textwidth}{!}{%
  \includegraphics{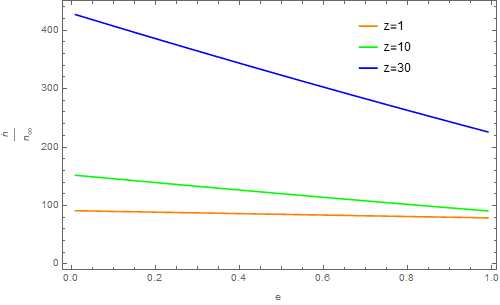}
}
\caption{Sample graphs of $\frac{\dot{n}}{n_{\infty}}$ as the function of $Q$ (left) and as the function of $e$ (right). The left parameters are chosen to be $e=0.1,\mu=0$, the right parameters are chosen to be $Q=0.5,\mu=0$.}
\label{Fig:Nra}
\end{figure}

We know that the outer horizon $r_{+}$ of a Reissner-Nordstr\"{o}m black hole requires that the mass of the black hole $M$ should be bigger than the charge $Q$. Say there are $N_0$ Fermions fall into the black hole, therefore the mass becomes $M+N_0m_0$ and the charge becomes $Q+N_0q$, where $M$ is the original mass and $Q$ is the original charge of the black hole. Then, the charge-mass ratio of a black hole $\frac{Q+N_0q}{M+N_0m_0}$ will change with the number of particles falling into the black hole. For those particles with $e\leq 1$ keeps falling into the black hole, the final charge-mass ratio of a black hole would inevitably turn into $e\leq1$. For those particles with $e> 1$, Wald's research \citep{Wald1974} shows that the accretion process, which destroys a black hole's event horizon, will be stopped. As is pointed by Ref.\citep{Wald1974} that the particles with $e> 1$ will not be captured by the extreme Reissner-Nordstr\"{o}m black hole $Q=M$. This is because that in the case of $e>1$, the electrostatic repulsion becomes greater than the gravitational attraction and the particle will be rejected. One can shooting it in with high velocity. But by doing so one increases the particle's energy. Until the energy is raised to $E>q_0$ which cannot violate the relationship $Q\leq M$, particles are to enter the black hole. Therefore, the naked singularity can never happened. To keep the horizon existing, we obtain the maximum number $N_{m}$ for the black hole accretion Fermions, which is
\begin{equation}
N_m=\left\{
    \begin{array}{ll}
        \frac{M-Q}{q-m_0}, & \hbox{for $q>m_0$;} \\
       \infty, & \hbox{for $q\leq m_0$.}
      \end{array}
    \right.
\end{equation}
This analysis shows that the accretion of Fermions onto a black hole would decrease as the black hole becomes more charged, which is shown in Fig.\ref{Fig:Nra}.

We plot the particle accretion rate $\frac{\dot{n}}{n_{\infty}}$ in Fig.\ref{Fig:Nra}. The left graph shows exactly $\dot{n}$ decrease as the charge $Q$ grows. The right graph also shows the accretion rate $\dot{n}$ decreases as $e$ increases. As $z$ increases, the accretion rate $\dot{n}$ increases. However, the accretion rate $\dot{n}$ depends little on the parameter $\mu$. The reason is that the parameter $\mu$ appears in the denominator of the integral function $\frac{1}{e^{z\varepsilon-\mu\beta}+1}$. The constant term $e^{\mu\beta}$ can be brought outside the integral and left the integral function to $\frac{1}{e^{z\varepsilon}+e^{\mu\beta}}$. The constant term $e^{\mu\beta}$ appears both in the numerator $\dot{n}$ and the denominator $n$. Thus, the effect of changing the parameter $\mu$ is equivalent to changing the $+1$ term in the distribution function (\ref{Fermidis}).

\begin{figure}
\resizebox{0.45\textwidth}{!}{%
  \includegraphics{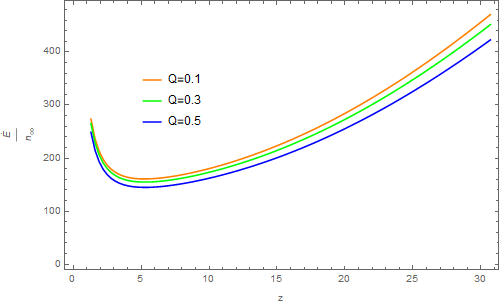}
}
\caption{Sample graphs of $\frac{\dot{E}}{n_{\infty}}$. The parameters are chosen to be $e=0.1,\mu=0$.}
\label{Fig:En}
\end{figure}

We plot the energy accretion rate $\frac{\dot{E}}{n_{\infty}}$ in Fig.\ref{Fig:En}. The energy accretion rate $\dot{E}$ is not monotonically dependent on $z$. As $z$ increases, $\dot{E}$ decreases to minimum and then increases. We also find that the accretion rate $\dot{E}$ decreases as $Q$ increases.

\subsection{energy density, radial and tangential pressures}
\begin{figure}
\resizebox{0.45\textwidth}{!}{%
  \includegraphics{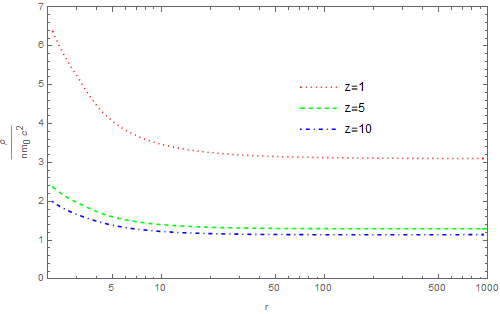}
}
\caption{Sample graphs of $\frac{\rho}{n m_0c^2}$. The parameters are chosen to be $Q=0.1,e=0.1,\mu=0$.}
\label{Fig:ende}
\end{figure}
\begin{figure}
\resizebox{0.45\textwidth}{!}{%
  \includegraphics{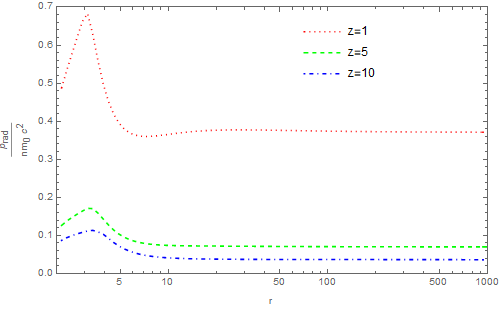}
}
\caption{Sample graphs of $\frac{p_{rad}}{n m_0c^2}$. The parameters are chosen to be $Q=0.1,e=0.1,\mu=0$.}
\label{Fig:prad}
\end{figure}
\begin{figure}
\resizebox{0.45\textwidth}{!}{%
  \includegraphics{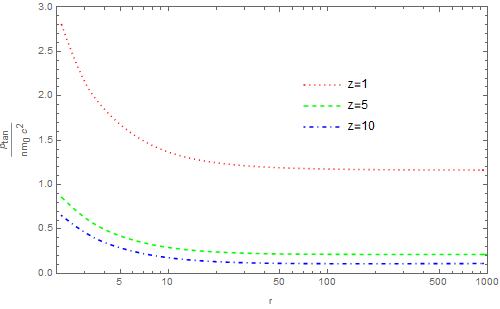}
}
\caption{Sample graphs of $\frac{p_{tan}}{n m_0c^2}$. The parameters are chosen to be $Q=0.1,e=0.1,\mu=0$.}
\label{Fig:ptan}
\end{figure}

\begin{figure}
\resizebox{0.45\textwidth}{!}{%
  \includegraphics{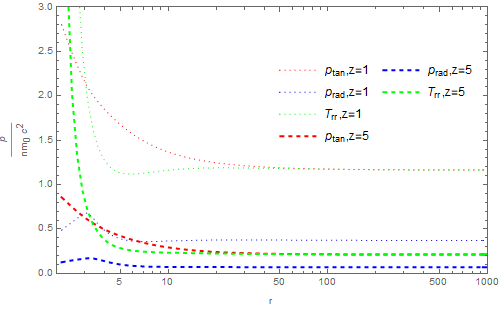}
}
\caption{A comparison among the radial, tangential pressure and $\frac{T_{rr}}{n m_0c^2}$ for two different values of $z$. }
\label{Fig:prpt}
\end{figure}

In Fig.\ref{Fig:ende}, we show the behavior of energy density $\frac{\rho}{n m_0c^2}$ as the function of $r$. Since we chose physical coordinates, one can more easily see how the physical quantity changes with the radius. The energy density $\frac{\rho}{n m_0c^2}$ changes similarly as the figure in Ref. \citep{Rioseco2017-2}. However, that seems no true for the radial pressure $\frac{p_{rad}}{n m_0c^2}$, see Fig.\ref{Fig:prad}. Fig.\ref{Fig:ptan} show that the tangential pressures $\frac{p_{tan}}{n m_0c^2}$ is consistent with Ref. \citep{Rioseco2017-2}. However, the radial pressure $p_{rad}$ is no longer equal to the tangential pressures $p_{tan}$ at infinity, see Fig.\ref{Fig:prpt}. As is shown in the previous section, $T^{r}{}_{r}$ equals to $p_{tan}$. And they are components of the energy-momentum of the isotropic perfect fluid for Fermi gas at infinity. Our numerical results show that the radial pressure $p_{rad}$ is always smaller than tangential pressures $p_{tan}$ even at infinity. As $z$ growth, the difference between $p_{rad}(r\rightarrow\infty)$ and $p_{tan}(r\rightarrow\infty)$ would get smaller and smaller. Only in the low-temperature limit $z\rightarrow \infty$, the contribution of null fluid disappears and the model can be treated as the ideal fluid accretion model at infinity.

\section{conclusions}
In this paper, we study the accretion of degenerate Fermi gas onto a Reissner-Nordstr\"{o}m black hole. The Boyer-Linquist coordinate is chosen to construct the accretion theory. We assume that the Fermi gas is in thermal equilibrium at infinity and described by Fermi-Dirac statistic. Each particle interacts with others only through the gravitational and electromagnetic field of the background and travels along the orbit described by the abbreviated action. We obtain the expression both for the absorption part and scattering part of the particle current density and the stress energy-momentum tensor. The expression of particle accretion rate and energy accretion rate are also obtained. At large distances, we analytically calculate the corresponding quantities. In the case of Maxwell-J\"{u}ttner distribution, they are consistent with the Ref. \citep{Rioseco2017-1,Cieslik2020} except $T^t{}_r$. At finite range, we numerically calculate three accretion rates, energy density, radial pressure and tangential pressure.

The Vlasov gas accretion model describes an anisotropic fluid consisting of two perfect-fluid components. One component is an isotropic perfect fluid of Fermi gas, and the other is a null fluid. When using Boyer-Lindquist coordinates, the contribution of the null fluid remains at infinity. At this point, the radial pressure is smaller than the tangential pressure. This difference decreases as the temperature decreases and eventually vanishes as the temperature approaches 0.

The results also show that the chemical potential had minimal effect on the results. The more charges of the black hole, the lower the accretion rates of a black hole. However, in the extremal case $Q=M$, the black hole would not accrete particles with $e>1$. The charge-mass ratio of a black hole would never be larger than 1. Thus, the naked singularity cannot form.

\section*{Acknowledgements}
This work is partially supported by the National Natural Science Foundation of China (NSFC U2031112) and the Scientific Research Foundation of Hunan University of Arts and Sciences (23BSQD237). We also acknowledge the science research grants from the China Manned Space Project with NO. CMS-CSST-2021-A06.

\bibliographystyle{elsarticle-harv} 
\bibliography{example}






\end{document}